\documentstyle[12pt]{article}
\begin{document}
\begin{center}
\begin{Large}
\section*{The Effects of Weak Gravitational Lensing on Determinations
of the Cosmology from Type Ia Supernov\ae}
\end{Large}
\end{center}

\noindent
Andrew J. Barber

\noindent
{\em Astronomy Centre, University of Sussex, Falmer, Brighton, BN1 9QJ,
U.K.}

\vspace{0.2in}
\subsection*{Abstract.}

Underlying cosmologies with deceleration parameter,
$q_0=-0.51^{+0.03}_{-0.24},$ may be interpreted as having $q_0=-0.55,$
on the basis of weak gravitational lensing effects.

\subsection*{1. Computing the Three-Dimensional Shear}

The algorithm used for determining the three-dimensional shear has
been described fully by Couchman, Barber \& Thomas (1998). Of
particular interest is the variable softening feature in the
algorithm, designed to distribute the mass of each particle within a
radial profile, which depends on its specific environment.

The algorithm has been applied to a number of different cosmological
$N$-body simulations available from the Hydra
consortium\footnote{http://coho.astro.uwo.ca/pub/data.html}; see, for
example, Barber, Thomas \& Couchman (1999).

\subsection*{2. The Effects of Weak Lensing on Determinations of the
Deceleration Parameter}

Both Riess et al. (1998) and Perlmutter et al. (1999), point to
cosmologies which are close to the $\Omega_M =0.3,$
$\Omega_{\Lambda}=0.7$ cosmological simulation (designated LCDM) I
have analyzed in terms of weak lensing. I shall therefore discuss the
dispersions in magnification and the impact on determinations of the
deceleration parameter, $q_0$, with regard to this assumed underlying
cosmology.

By considering Type Ia supernov\ae~sources at ten evenly spaced
redshifts, $z$, between $z=0$ and $z=1$, the resulting distributions
of the magnifications show the expected increasing dispersions with
$z,$ which could be interpreted as observational dispersions in the
peak magnitudes. They are equivalent to differences in distance
modulus, $m$, and these differences have been evaluated to correspond
to the $2\sigma$ range in the magnification distribution, denoted by
$\mu_{\mathrm low}$ and $\mu_{\mathrm high}.$

Riess et al. (1998) have recognized that weak lensing of high-redshift
Type Ia supernov\ae~can alter the observed magnitudes and quote the
findings of Wambsganss et al. (1997), i.e., that the light will be on
average demagnified by 1\% at $z=1$ in a universe with a
non-negligible cosmological constant. They state that the size of this
effect is negligible. Perlmutter et al. (1999) assume that the effects
of magnification or demagnification will average out, and that the
most over-dense lines of sight should be rare for their set of 42
high-redshift Type Ia supernov\ae. However, they note that the average
(de)amplification bias from integration of the probability
distributions is less than 1\% for redshifts of $z\leq1.$ My work has
shown that, in the LCDM cosmology, a source at $z=1$ will have on
average (de)amplification bias of 3.4\% $(\mu_{\mathrm peak}=0.966)$,
and 1.3\% at $z=0.5$. Whilst these effects are still rather small,
there is a significant probability of observing highly magnified Type
Ia supernov\ae; $97\frac{1}{2}\%$ of all lines of sight display a
range of magnifications up to 1.191 at $z=1,$ and this range introduces
an approximate $2\sigma$ (skewed) dispersion of 0.252 magnitudes.

On the evidence of the data, an underlying cosmology with $q_0 = -0.51
^{=0.03}_{-0.24}$ may be interpreted as having $q_0 =-0.55,$ from the
use of perfect standard candles (without intrinsic dispersion),
arising purely from the effects of weak lensing. This dispersion in
$q_0$ is somewhat larger than that found by Wambsganss et al. (1997)
based on a cosmology with $\Omega_m = 0.4,$ $\Omega_{\Lambda}= 0.6$,
because of the broader magnification distribution at $z=1$. Fluke (1999) has analyzed his data differently to obtain
dispersions in $q_0$. His much larger dispersions compared with
mine, come primarily from the much higher values of $\mu_{\mathrm
high}$ in his results.

\subsection*{Acknowledgments} 

I acknowledge the dirct help of Peter A. Thomas of the University of
Sussex and H. M. P. Couchman of McMaster University. The Starlink
minor node at the University of Sussex has been used in the
preparation of this paper.

\subsection*{References}

\baselineskip 0.41cm 
\begin{trivlist} 
\vspace{0.1in}
\item{Barber A. J., Thomas P. A. \& Couchman H. M. P., 1999, astro-ph 9901143.}
\vspace{0.1in}
\item{Couchman H. M. P., Barber A. J. and Thomas P. A., 1998, astro-ph
9810063.}
\vspace{0.1in}
\item{Fluke C. J., 1999, Ph. D. Thesis, University of Melbourne.}
\vspace{0.1in}
\item{Perlmutter S., Aldering G., Goldhaber G., Knop R. A., Nugent P.,
Castro P. G., Deustua S., Fabbro S., Goobar A., Groom D. E., Hook
I. M., Kim A. G., Kim M. Y., Lee J. C., Nunes N. J., Pain R.,
Pennypacker C. R., Quimby R., Lidman C., Ellis R. S., Irwin M.,
McMahon R. G., Ruiz-Lapuente P., Walton N., Schaefer B., Boyle B. J.,
Filippenko A.  V., Matheson T., Fruchter A. S., Panagia N., Newberg H.
J. M., Couch W. J., and Project T. S. C., 1999, Ap. J., 517, 565.}
\vspace{0.1in}
\item{Riess A. G., Filippenko A. V., Challis P., Clocchiatti A.,
Diercks A., Garnavich P. M., Gilliland R. L., Hogan C. J., Jha S.,
Kirshner R. P., Leibundgut B., Phillips M. M., Reiss D., Schmidt
B. P., Schommer R. A., Smith R. C., Spyromilio J., Stubbs C., Suntzeff
N. B. and Tonry J., 1998, A. J., 116, 1009.}
\vspace{0.1in}
\item{Wambsganss J., Cen R., Xu G. and Ostriker
J. P., 1997, Ap. J., 475, L81.}

\end{trivlist}
\begin{description} 
\item 
\end{description} 

\end{document}